\documentclass[twocolumn,showpacs,preprintnumbers,amsmath,amssymb,prl]{revtex4}

\usepackage{epsfig}
\usepackage{dcolumn}
\usepackage{bm}
\usepackage{latexsym}

\begin{document}

\title{Resonant interaction of molecular vibrations and surface plasmon polaritons: The weak coupling regime}

\author{S. Kalusniak, S. Sadofev, F. Henneberger}
\affiliation{ Institut f\"ur Physik, Humboldt Universit\"at zu Berlin, Newtonstr.15, 12489 Berlin, Germany}

\date{\today}

\begin{abstract}
Adjusting the free-electron concentration, the surface plasmon frequency of the semiconductor ZnOGa is tuned into resonance with the molecular vibrations of the n-alkane tetracontane. Closed molecular films deposited on the semiconductor's surface in the monolayer regime generate distinct signatures in total-attenuated-reflection spectra at the frequencies of the symmetric and asymmetric stretching vibrations of the CH$_{2}$ group. Their line shape undergoes profound changes from absorptive to dispersive and even anti-resonance behavior when moving along the surface- plasmon dispersion by the angel of incidence. We demonstrate that this line shape diversity results from a phase-sensitive perturbation of the surface-plasmon-polariton generation at the molecule/metal interface.
\end{abstract}



\maketitle

\begin{figure}
\includegraphics[width = 7.5 cm]{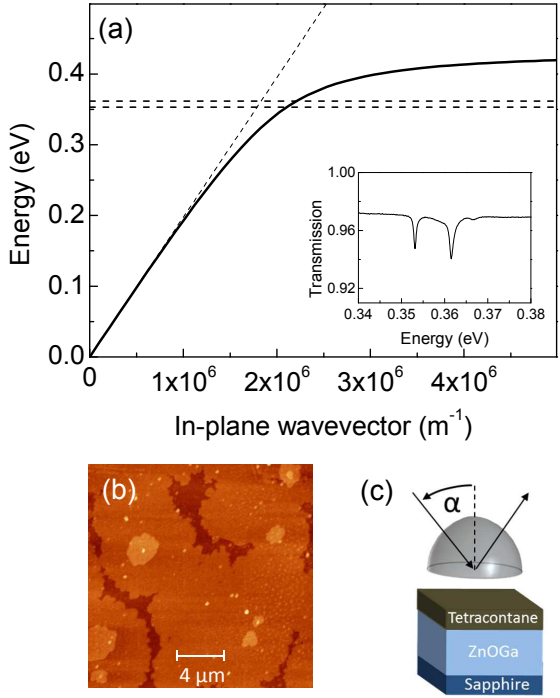}
\caption{(a) SPP dispersion of the sole air/ZnOGa interface and positions of the symmetric and asymmetric CH$_{2}$ vibrational resonances of C40. The dispersion is derived from $[1+\epsilon_p(\omega)] c^{2}k_\parallel^{2}=\omega^{2}\epsilon_p(\omega)$  with $\epsilon_p(\omega)=\epsilon_{bp}-\omega_p^2/(\omega^2-2 i \omega \gamma_p)$. Plasma frequency $\hbar \omega_p = 0.94$ eV and electron scattering rate $\hbar\gamma_p = 63$ meV at the chosen doping level are taken from free-space absorption and Hall measurements on single ZnOGa films \cite{Sadofev13}. The surface plasmon frequency is $\hbar\omega_{sp}=\hbar\omega_p/\sqrt{1+\epsilon_{bp}}=$ 0.43 eV. Dotted: photon line. Inset: Transmission spectra of C40 on sapphire recorded for reference. The dashed vertical lines in the main panel represent the two transmission minima. (b) Atomic force image of the C40 layer with almost upright-standing molecules on the ZnOGa surface. (c) Schematics of the ATR measurements with the Si hemisphere. The in-plane wavevector of the dispersion in (a) is set by the angle of incidence $\alpha$ through $k_\parallel=\sqrt{\epsilon_{Si}} (\omega/c) \sin \alpha$. For more details of the set-up, see Ref. \cite{Kalus14}.}
\label{fig:Fig1}
\end{figure}

\begin{figure}
\includegraphics[width=8.0 cm]{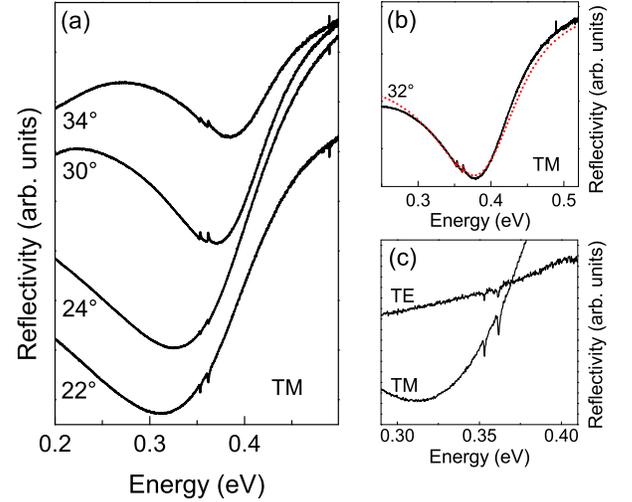}
\caption{ATR spectra of the C40/ZnOGa specimen described in the text. (a) TM polarization for different angles of incidence (denoted at the curves). (b) Example for agreement between experimental (full curve) and calculated (dotted curve) spectra in TM polarization. (c) Comparison of TE and TM polarization at maximum molecular signal ($\alpha=22^\circ$). For parameters used in the transfer matrix computations, see caption of Fig. \ref{fig:Fig1} (ZnOGa) and text. In (a) and (c) spectra are vertically shifted to increase visibility of the molecular features.}
\label{fig:Fig2}
\end{figure}

\begin{figure}
\includegraphics*[width = 8.0 cm]{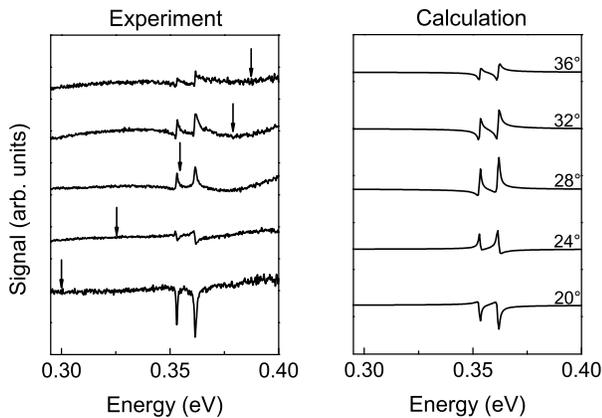}
\caption{ATR spectra in vicinity of the vibrational resonances. The baseline (zero signal) is related to the background signal obtained by (i) extrapolating the spectra from sufficiently far distances continuously through the resonance region for the experimental data and (ii) by switching off the resonance contribution ($\epsilon_m=\epsilon_{bm}$) in the calculations. The arrows mark the position of the broad ATR minimum of the SPP feature.}
\label{fig:Fig3}
\end{figure}

\begin{figure}
\includegraphics*[width = 8.0 cm]{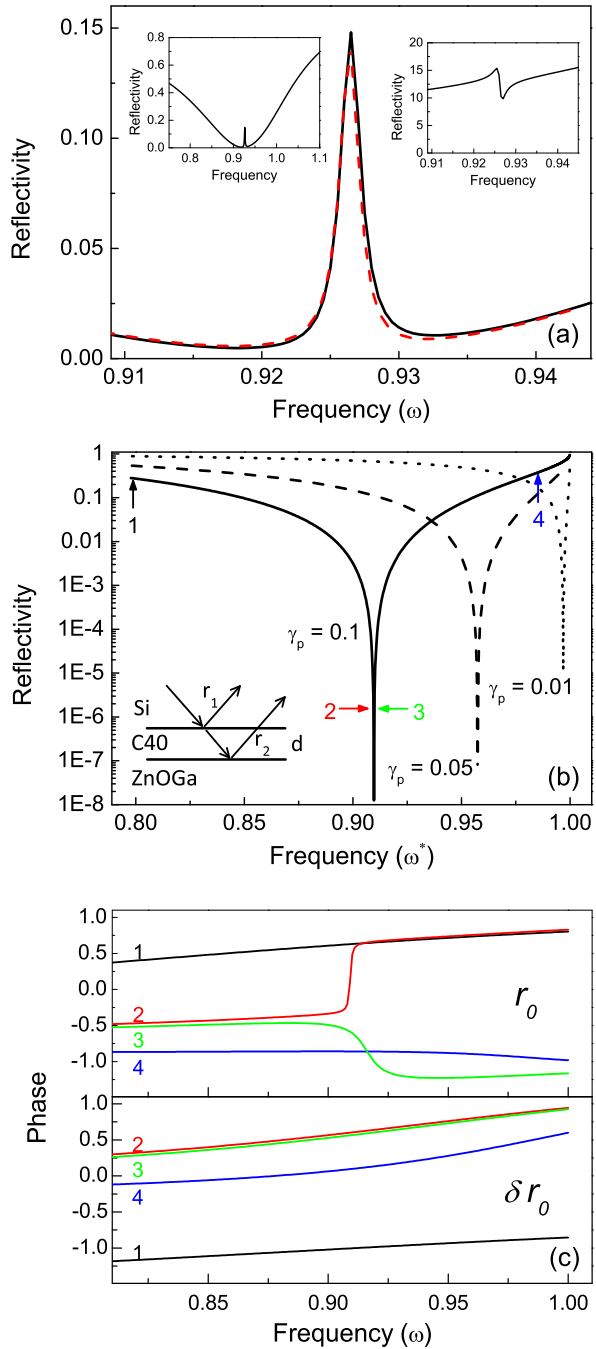}
\caption{ATR line shape of the molecular signal in the weak coupling regime. All frequencies are measured in units of $\omega_{sp}^*$.
(a) Resonance case: Inverted absorptive line shape. Full curve: Exact slab reflectivity $|r|^2$. Dashed curve: Account of the molecular resonance only in $r_2$. For other detunings, the agreement is similarly perfect. Left inset: Overview spectrum. Right inset: $|r_2|^2$ in vicinity of the molecular transition. (b) Depth of the reflectivity minimum in $|r_0|^2$ as function of the SPP frequency. Inset: Configuration analyzed. (c) Phases of $r_0$ and $\delta r_0$ versus frequency $\omega$ at the $\omega^*$ marked by arrows in (b).  Parameters:  $\omega_L^2-\omega_T^2=10^{-4},2\gamma_m=10^{-3}$ [for (a) only], $\gamma_p=0.1$ [and $\gamma_p=0.05, 0.01$ in (b)], all in units of the half-space surface-plasmon frequency $\omega_p/\sqrt{\epsilon_{bp}+\epsilon_{bm}}$, $d=0.05~ (2\pi c/ \omega_{sp})$.}
\label{fig:Fig4}
\end{figure}

The coupling of surface plasmons and molecular excitations is an extensively studied subject, both from a practical as well as fundamental point of view. In particular, the direct infrared detection of molecular vibrations has profited dramatically from the utilization of a plasmon-related field enhancement. Techniques such as surface-enhanced Raman scattering \cite{Moskovits85,Nie97} or infrared absorption \cite{Wang07, Aroca06} have enabled sensitivity increases of several orders of magnitude. However, despite this progress, the physics underlying these observations is not always fully understood \cite{Hart80,Hatta84,Osawa01}. Previous work in this context can be roughly divided in two groups. Already decades ago, so-called resonance transition layers deposited on the surface of traditional metals have been investigated \cite{Agran74, AbZhi82}. As the surface plasmon resonance of these metals is deep in the ultraviolet spectral range, the coupling with the vibrational excitations is strongly off-resonant. Only if the vibrational transition is so strong that its permittivity contribution becomes negative as well, a significant coupling occurs and manifests by a splitting and broadening of the attenuated-total-reflection (ATR) features as compared to the uncovered metal \cite{Layer}. More recently, enabled by advances in nanotechnology, the coupling of molecular vibrations to localized surface plasmons of metal nanoaggregates or lithographically defined antennas has been studied \cite{Neubrech08,Giannini11,Frontiera12}. Here, the plasmonic resonances of the metal entities can be indeed tuned into resonance with the molecular transitions. As a result, Fano-type features at the vibrational frequencies have been observed indicating interference between molecular and surface plasmon excitations.  Similar observations were made using a metamaterial built from coupled split-ring resonators \cite{Pryce10}.\\
In this Letter, we return to an extended geometry where surface plasmon polaritons (SPPs) propagate at a planar interface. However, instead of using a metal, our study is based on a heavily doped semiconductor - ZnOGa - where the surface plasmon frequency can be adjusted almost arbitrarily from the mid to near infrared range through the free-electron concentration \cite{Kalus14}. Unlike previous work \cite{Hart80,Hatta84,Osawa01,Agran74, AbZhi82}, we find distinct signatures of resonance coupling also for weak vibrational transitions. As our setting is well-defined and geometrically simple, we are able to isolate the clean electromagnetic response. Chemical interactions often invoked and debated when interpreting surface-enhanced spectroscopy data can be safely excluded here.  A further attractive feature of the setting is that, unlike localized surface plasmons \cite{Neubrech08,Giannini11,Frontiera12,Pryce10},  vibrational energy can be coherently transported along the surface.\\
The ZnOGa films are grown on sapphire substrates by molecular beam epitaxy as described previously \cite{Sadofev13}. The Ga doping level is chosen such that the air/ZnO surface plasmon frequency exceeds the frequency of the symmetric (s) and asymmetric (a) stretching vibrations of the CH$_{2}$ group of the n-alkane tetracontane (C40) less than the plasmonic broadening [Fig.\ref{fig:Fig1}(a)]. For ensuring monolayer control, the molecules are attached to the ZnOGa surface by organic molecular beam deposition. For the present study, a sample with a closed C40 layer of two-monolayer average thickness is selected.  As demonstrated by the atomic-force-microscopy image in Fig.\ref{fig:Fig1}(b), the area fraction with one- and three-monolayer coverage is less than 10 \%. ATR spectra are recorded by a Bruker IFS66v/s vacuum Fourier transform spectrometer in Otto configuration with a polished silicon hemisphere as coupling medium [Fig.\ref{fig:Fig1}(c)].\\
Figure \ref{fig:Fig2} summarizes ATR overview spectra. Molecular signatures are clearly present. For transverse magnetic (TM) polarization, they appear superimposed to the prominent band associated with the SPP generation. Unlike the case of metal nanostructures \cite{Neubrech08,Giannini11}, where the size has to be properly adjusted in order to fit the localized plasmonic resonance to the vibrational transition, the ATR minimum can be tuned in the present configuration continuously across the molecular resonances for a given sample via the in-plane vector $k_\parallel$ defined experimentally by the angle of incidence $\alpha$. The experimental spectra are well reproduced by transfer matrix calculations [Fig. \ref{fig:Fig2}(b)] accounting for the silicon hemisphere as semi-infinite medium ($\epsilon_{Si}=11.8$), the air gap ($\epsilon=1$), the molecular layer, the ZnOGa film, as well as the sapphire substrate again as semi-infinite medium ($\epsilon =2.89$). While the thicknesses of the molecular layer (10 nm) and the ZnOGa film (400 nm) are precisely known, the width of the air gap is fine-adjusted such that experimental and calculated SPP minima best match. As justified previously \cite{Sadofev13}, the permittivity of ZnOGa is described by the Drude dielectric function with a high-frequency background contribution $\epsilon_{bp}=3.7$ of pure ZnO. The vibrational transitions are treated as Lorentzian resonances
\begin{eqnarray} \label{DFM}
\epsilon_m(\omega)=\epsilon_{bm}(1+\sum_{j=s,a}\frac{\omega^2_{Lj}-\omega^2_{Tj}}{\omega^2_{Tj}-\omega^2-2i\omega\gamma_j})
\end{eqnarray}
where $\omega_{Tj}$ and $\omega_{Lj}$ are the transverse and longitudinal frequencies, respectively, $\gamma_j$ are the vibrational damping rates, and $\epsilon_{bm}$ is the background permittivity of the C40 layer. The positions of the SPP features are low-frequency shifted with respect to the half-space air/ZnOGa dispersion in Fig. \ref{fig:Fig1}(a), both by the finite air gap as well as by presence of the molecular film. As detailed below, the latter shift is almost completely due to the layer's background permittivity $\epsilon_{bm}$, as the resonance contribution of the vibrational transitions is small [$(\omega_{Lj}-\omega_{Tj} )/ 2 \gamma_j  < 0.3$]. A weaker molecular signal is also observable in transverse electric (TE) polarization with otherwise structureless spectra [Fig. \ref{fig:Fig2}(c)].\\
More information about the molecular resonances is attained from the zoomed plots of Fig. \ref{fig:Fig3} where the signal from the off-resonant background is subtracted. There are various points that deserve attention. First, the molecular parameters  [$\hbar\omega_{T}=0.3531$ eV (s), 0.3615 eV (a), $\hbar\omega_{L}-\hbar\omega_{T}$=0.16 meV (s), 0.25 meV (a), $\gamma_{s}=0.34 $ meV, $\gamma_{a}=0.40 $ meV, $\epsilon_{bm}=2.19$] assumed for reproducing the experimental spectra for ZnOGa are identical in TE and TM polarization as well as with those of the sapphire reference sample. That is, the coupling to the SPPs in TM configuration is indeed weak as it does not modify the molecular resonances. The parameters agree also reasonably well with literature data for C40 \cite{C40}. Second, the small broadening ($\gamma_m <$ 1 meV) of both vibrations demonstrates that no significant disorder or extra dissipation is introduced, neither by chemical interactions with the ZnOGa substrate nor by coupling to the SPPs. Third, while the vibrational transitions show always up as dips in the TE reflectivity [like in the example of Fig. \ref{fig:Fig2}(c)],  their  line shape as well as strength undergo distinct changes in dependence on the detuning from the SPP feature in TM polarization. This observation is in marked contrast to previous ATR studies on molecules deposited on Ag films where merely absorptive shapes were found \cite{Hart80,Hatta84,Osawa01}. Obviously, a novel phase-controlled coupling of the surface SPPs and the molecular vibrations occurs under resonance conditions which is analyzed in detail in what follows.\\
For simplicity we ignore the air gap as it merely determines the absolute position of the surface plasmon frequency $\omega_{sp}^*=\lim_{k_\parallel\rightarrow \infty}\omega(k_\parallel)$ in the specific ATR setting, whereas the detuning of the molecular transition with respect to that frequency is the essential parameter. Then, the amplitude reflectivity of the Si/C40/ZnOGa configuration is given by the standard formula $r= (r_1+r_2 \text e^{2i\varphi})/(1+r_1 r_2 \text e^{2i\varphi})$ for a slab of thickness $d$
where $r_1$ and $r_2$ are the reflectivity at the first and second interface, respectively, and $\varphi=\sqrt{(\omega/c)^2\epsilon_m-k^2_\parallel}d$ is the single-path (complex) phase shift [inset Fig. \ref{fig:Fig4}(b)]. In ATR, the dominant contribution of the molecular transition in TM polarization originates from $r_2$. The molecule-induced phase shift is negligible for the ultra-thin layers under study and $|r_1|=1$, while $r_2 >1$ because of the plasmonic enhancement [Fig. \ref{fig:Fig4}(a)]. If the molecular reflectivity contribution $\delta r$ is sufficiently small, the resultant reflection change can be approximated by $\delta R= |r_0+\delta r|^2- |r_0|^2\approx 2 \text{Re}~( r_0 \overline{\delta r})$ with $r_0$ as the slab reflectivity without the molecular resonance ($\epsilon_m=\epsilon_{bm}$). The leading part of $\delta r$ can be expressed by $\delta r =\delta r_0(\epsilon_m-\epsilon_{bm})$ with the derivative $\delta r_0=( \partial r/\partial r_2)(\partial r_2/\partial \epsilon_m)|_{\epsilon_m=\epsilon_{bm}}$ providing
\begin{equation} \label{DR}
\begin{split}
\delta R =  &2 |r_0 \delta r_0| \\
&\left[ \cos(\delta \phi)~ \text{Re}(\epsilon_m-\epsilon_{bm})
+\sin(\delta \phi) ~ \text{Im}~ \epsilon_m \right] \\
\end{split}
\end{equation}
where $\delta \phi=\arg (r_0) - \arg (\delta r_0)$. This expression directly demonstrates that the molecule-induced signal is determined by the characteristics of the sole plasmonic response, namely by the reflectivity itself as well as the reflectivity change per permittivity unit. In particular, the phases of these quantities control wether the line shape is absorptive or dispersive and whether it has positive or negative sign. By inserting the explicit expression for $r_1$ and $r_2$ from the Fresnel equations in TM polarization with the respective dielectric functions, the frequency and angle dependence of $\delta R$ is obtained. The angle $\alpha$, again specific of the concrete ATR setting, is eliminated by replacing it by the frequency $\omega^*$ of the corresponding minimum in the reflection spectrum $R_0=|r_0|^2$, i.e., $\omega^*$ defines the frequency at which the SPP is excited within its dispersion curve, while $\omega$ scans the spectrum at that selected excitation. Normalizing $\omega$ and $\omega^*$ to $\omega^*_{sp}$, the only remaining free parameter in the plasmonic response is the broadening $\gamma_p$. \\
\begin{figure}
\includegraphics*[width = 8.0 cm]{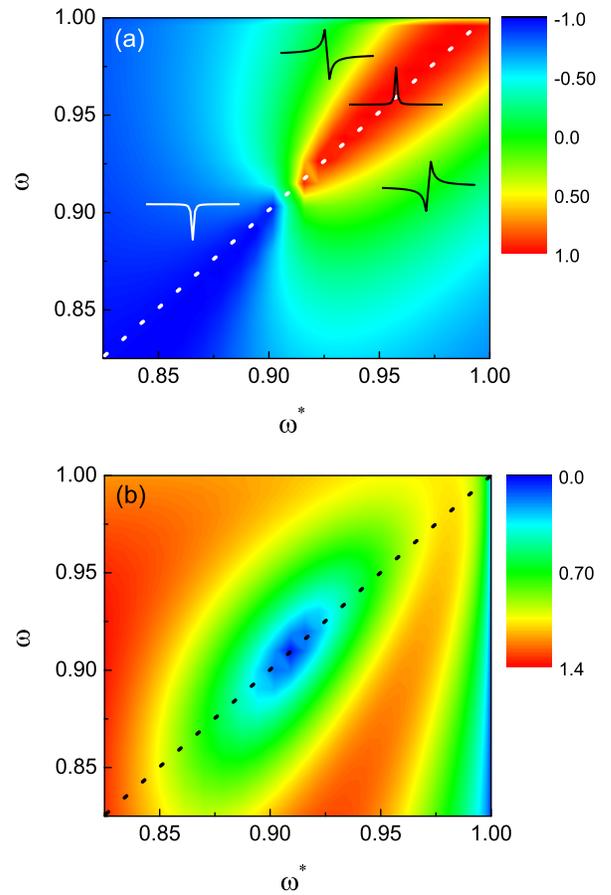}
\caption{Plasmonic environment: Contour plot in the $\omega\omega^*$-plane of (a) the phase term  $\sin(\delta \phi)$ and (b) the amplitude $2|r_0\delta r_0|$ term in Eq. \ref{DR}. Parameters as in Fig. \ref{fig:Fig4}. The inserted curves depict symbolically the line shape of the molecular transition at representative locations.}
\label{fig:Fig5}
\end{figure}
Figure \ref{fig:Fig4} and \ref{fig:Fig5} summarize the scenario for parameters typical of the weak coupling regime and similar to the experiment. The depth of the minimum of $R_0$ exhibits in turn a distinct minimum as a function of $\omega^*$ [Fig. \ref{fig:Fig4}(b)]. Here, destructive interference between the waves reflected from the first and second interface occurs so that the nominator of $r_0$ approaches zero. The position of the minimum ($\omega^*_{min}$) depends on the plasmonic damping and moves towards $\omega^*_{sp}$ when $\gamma_p$ decreases. The phase of $r_0$ behaves discontinuous at $\omega^*_{min}$  [Fig. \ref{fig:Fig4}(c)]. While it is a monotonically increasing function of $\omega$ for $\omega^* < \omega^*_{min}$, there is a switch to monotonic decrease for $\omega^* > \omega^*_{min}$. As the phase of $\delta r_0$ exhibits no such peculiarity [Fig. \ref{fig:Fig4}(c)], it is that discontinuity what causes the line shape change of the molecular signal. Figure \ref{fig:Fig5}(a) and (b) depict contour plots in the $\omega\omega^*$-plane of the $\sin(\delta \phi)$- and $2|r_0\delta r_0|$-term, respectively, in Eq. (\ref{DR}). From those plots it can be directly inferred in what plasmonic environment the molecular transition has to be placed in order to realize a certain line shape: The molecular resonance frequency defines the position on the $\omega$-axis, while the location on the $\omega^*$-axis measures the detuning relative to the surface plasmon frequency.  If that detuning exceeds markedly the plasmonic damping, the molecular signal is always absorptive. However, when the molecular transition is located closer to $\omega^*_{sp}$, the line-shape changes to dispersive and then to the corresponding anti-resonance features when $\omega^*$ crosses the molecular position from low to high frequencies. Note also that the maximum signal does not simply appear for $\omega^*=\omega$. All these deductions are fully consistent with the experimental findings.\\

In conclusion, both our experimental study and the theoretical analysis reveal how a weak Lorentzian resonance affects the SPPs of an adjacent metal. Its absorption and refraction contribution modifies the generation condition for SPPs at the interface and is so imprinted in the spectral response. Under resonant coupling $(\omega_T \sim \omega_{sp})$, this gives rise to a diversity of line shapes depending on the specific phase adjustment. A phase-controlled variation between absorptive and dispersive features is quite common in optics. Moreover, a quasi-continuous SPP spectrum is not a requirement. Also for the unrealistic case $(\gamma_p \approx \gamma)$, the calculations predict complex line-shape features. Spectrally broad SPP states make merely the experimental observation more easy. We avoid thus the term "Fano-type" in this context. The weak enhancement of the molecular signal is fully consistent with a strict planar geometry. Similarly as demonstrated for metals, surface roughness or artificial nanostructuring will substantially increase the enhancement factor. For practical applications, semiconductors have the advantage of a negative but small permittivity in the relevant spectral range, favorable for the formation of strong local surface fields.

This work was supported by the Deutsche Forschungsgemeinschaft in the frame of SFB 951 (HIOS). The authors thank Maurizio Roczen for growing the C40 layer and Moritz Eyer for the atomic-force-microscopy measurements.


\begin{references}
\bibitem{Moskovits85}
M. Moskovits, Phys. Mod. Phys. {\bf 57}, 783 (1985).
\bibitem{Nie97}
S. M. Nie and S. R. Emory, Science  {\bf 275}, 1102 (1997).
\bibitem{Wang07}
H. Wang, J. Kundu, and N. J. Halas, Angew. Chem. 46, 9040 (2007).
\bibitem{Aroca06}
R. Aroca, Surface-Enhanced Vibrational Spectroscopy (John Wiley \& Sons, Ltd., New York, 2006).
\bibitem{Hart80}
A. Hartstein, J.R. Kirtley, and J. C. Tsang, Phys. Rev. Lett. {\bf 45}, 201 (1980).
\bibitem{Hatta84}
A. Hatta, Y.Suzuki, and W. Sueteka,  Appl. Phys. A {\bf 35}, 135 (1984).
\bibitem{Osawa01}
M. Osawa, Topics Appl.- Phys. {\bf 81} 163 (2001).
\bibitem{Agran74}
V. M. Agranovich and A. G. Malshukov, Optics Commun.  {\bf 11}, 169 (1974).
\bibitem{AbZhi82}
F. Abeles and T. Lopez-Rios, Surface polaritons at metal surfaces and interfaces, in Surface Polaritons (eds Agranovich, V.M. and Mills, D.L.) (North-Holland, Amsterdam, 1982).
G. N. Zhizhin and V. A. Yakovlev, Resonance of transition layer excitations with surface polaritons, ibid.,  and references therein.
\bibitem{Layer}
In that case, the vibrational resonance itself is capable of forming SPPs at its interface to air provided the layer is thick enough.
\bibitem{Neubrech08}
F. Neubrech, A. Pucci, T. W. Cornelius, S. Karim, A. Garc\'{i}a-Etxarri, and J. Aizpurua, Phys. Rev. Lett. {\bf 101}, 157403 (2008).
\bibitem{Giannini11}
V. Giannini, Y. Francescato, H. Amrania, C. C. Phillips, and S. A. Maier, NanoLett. {\bf 11}, 2835 (2011).
\bibitem{Frontiera12}
R. R. Frontiera, N. L. Gruenke, and R. P. Van Duyne, NanoLett. {\bf 12}, 5989 (2012).
\bibitem{Pryce10}
I. M. Pryce, K.  Aydin, Y. A. Kelaita, R. M. Briggs, and H. A. Atwater, NanoLett. {\bf 10}, 4222 (2010).
\bibitem{Kalus14}
S. Kalusniak, S. Sadofev, and F. Henneberger, Phys. Rev. Lett.  {\bf 112}, 137401  (2014).
\bibitem{Sadofev13}
S. Sadofev, S. Kalusniak, P. Sch\"afer, and F. Henneberger,  Appl. Phys. Lett.  {\bf 102}, 181905 (2013).
\bibitem{C40}
On KBr, the resonances ($\omega_T$) are blue-shifted by less than  3 cm$^{-1}$ [W. W. Duley, V. I Grishko, and G. Pinho, ApJ {\bf 642}, 966 (2006)]. The refractive index of C40 specified by different suppliers varies from 1.46-1.49 which is consistent with our $\epsilon_{bm}$-value. The $\omega_L$-frequencies determining the oscillator strength of the transitions are effective values, as the evanescent field polarization averages over the orientation of the vibrational dipole moment in the ATR geometry.

\end{references}
\end{document}